# Designer Quantum States in Magnetic Topological Insulator Multilayers

Deyi Zhuo[1], Han Tay[1], and Cui-Zu Chang[1]

[1]Department of Physics, The Pennsylvania State University, University Park, PA 16802, USA

Corresponding author: cxc955@psu.edu (C.-Z. C.).

**Abstract: Magnetic topological insulators (TIs) provide a highly tunable platform for engineering quantum states that emerge from the interplay between topology and magnetism. In this review article, we summarize experimental progress over the past decade in designing magnetic TI multilayers by molecular beam epitaxy (MBE). By treating magnetically doped and undoped TI layers as "topological Legos," we discuss how layer thickness, magnetic doping, heterostructure architecture, and stacking sequence can be used to control magnetic exchange gaps, interlayer coupling, and the Chern number $C$ with atomic-layer precision. We first briefly review the realization of the $C = 1$ quantum anomalous Hall (QAH) effect in uniformly Cr-doped $(Bi,Sb)_2Te_3$ films in 2013 and uniformly V-doped $(Bi,Sb)_2Te_3$ films in 2015. We then discuss how Cr-doped and undoped $(Bi,Sb)_2Te_3$ layers can be combined to realize the $C = 1$ QAH effect in magnetically modulation-doped trilayers, including its extension into the three-dimensional (3D) regime. Next, we review the development of high-$C$ QAH states, engineered plateau phase transitions, mesoscopic QAH devices, and electrical switching of chiral edge-current chirality. Finally, we discuss the realizations of axion insulator and $C = 1/2$ parity anomaly states in asymmetric magnetic TI trilayers. These advances establish magnetic TI multilayers as a versatile materials platform for creating new designer quantum states, including synthetic Weyl semimetal and QAH metal phases, and for probing the topological magnetoelectric effect in thick axion insulators and 3D QAH insulators.**

## 1. Introduction

The ability to engineer and control quantum states of matter represents both a central challenge in condensed matter physics and a strategic opportunity for next-generation quantum technologies. Among the most prominent platforms are topological insulators (TIs), a class of quantum materials that are electrically insulating in the bulk but host conducting states at their edges or surfaces. These boundary states arise from the nontrivial topology of the bulk electronic band structure and are protected by time-reversal symmetry (TRS) [1-3]. Introducing magnetism into a TI breaks TRS and gives rise to entirely new quantum phases [4-10], including the quantum anomalous Hall (QAH) effect (Figs. 1, 2b, 3, and 4) [11-15] and the axion insulator state (Figs. 2c and 7a) [16,17]. These emergent quantum states highlight the potential of magnetic TIs as a versatile platform for engineering topological transport and electromagnetic responses in designed heterostructures.

The QAH effect was theoretically proposed by Duncan Haldane in 1988 (ref. [15]) and experimentally realized in 2013 in a molecular beam epitaxy (MBE)-grown magnetically doped TI film, specifically 5 quintuple-layer (QL) Cr-doped $(Bi,Sb)_2Te_3$ (Fig. 1b) [12]. This breakthrough was featured in the scientific background of the 2016 Physics Nobel Prize honoring Duncan Haldane [18]. In 2015, a high-precision QAH state was realized in a 4 QL V-doped $(Bi,Sb)_2Te_3$ film (Fig. 1c) [19-21]. The QAH effect in magnetically doped TI systems has since been independently confirmed by research groups worldwide [13,14,21-32]. In addition to MBE-grown magnetically doped TI films and heterostructures, zero-magnetic-field integer and fractional QAH effects have also been realized in manually exfoliated intrinsic magnetic TI $MnBi_2Te_4$ (refs. [33-36]), rhombohedral-stacked multilayer graphene [37-39], and moiré materials, including twisted bilayer graphene [40], AB stacked $MoTe_2/WSe_2$ (ref. [41]), and twisted bilayer $MoTe_2$ (refs. [42,43]). The resistance-free chiral edge channel (CEC) of the QAH state is predicted to enable next-generation ultralow-power electronic and spintronic devices, as well as topological quantum computation [5,7-10].

A key advance in this field is the use of MBE to construct magnetic TI heterostructures with atomic-layer precision. By treating magnetically doped and undoped $(Bi,Sb)_2Te_3$ layers as "topological Legos", one can systematically control layer thickness, magnetic doping, and stacking sequence, thereby tuning magnetic exchange gap sizes and interlayer coupling strength (Fig. 2a). In magnetic TI trilayers, the relative magnetization alignment of the top and bottom magnetic TI layers determines the resulting topological phase. Parallel alignment produces the QAH state (Figs. 2b and 3a), characterized by a CEC (refs. [12-14,19-32,44]). In contrast, antiparallel alignment yields the axion insulator state (Figs. 2c and 7a), in which all surfaces are gapped and a quantized topological magnetoelectric (TME) response emerges [16,17]. Experimentally, such configurations can be realized by combining magnetic TI layers with different coercive fields, such as Cr- and V-doped $(Bi,Sb)_2Te_3$ (refs. [12,19,28]).

Extending this "topological Legos" approach to asymmetric and multilayer structures enables access to a broader class of quantum states. For example, a Cr-doped $(Bi,Sb)_2Te_3/(Bi,Sb)_2Te_3$ bilayer heterostructure can selectively gap only one surface Dirac cone, realizing a parity anomaly state (Fig. 2d). More generally, stacking multiple Cr-doped and undoped $(Bi,Sb)_2Te_3$ layers allows independent tuning of quantum confinement, exchange coupling, and hybridization (Fig. 2e). This capability enables deterministic control of the Chern number $C$, as demonstrated in multilayer structures such as [3 QL Cr-doped $(Bi,Sb)_2Te_3$/4 QL $(Bi,Sb)_2Te_3]_m$/3 QL Cr-doped $(Bi,Sb)_2Te_3$, where high-$C$ QAH states up to $C = 5$ were realized (Fig. 4) [4,44]. Here, $m$ is an integer. These advances establish magnetic TI multilayers as a powerful and flexible platform for engineering designer quantum phases. Through atomic-scale control of heterostructure design, it becomes possible to systematically explore and manipulate topological states, including high-$C$ QAH phases, axion insulators, and parity anomaly states, as well as more complex quantum phenomena emerging from the interplay between topology and magnetism.

## 2. The QAH effect

### 2.1 *C* = 1 QAH effect in magnetic TI trilayers

As noted above, the $C = 1$ QAH effect has been realized in uniformly Cr- and V-doped $(Bi,Sb)_2Te_3$ films. However, in these uniformly magnetically doped TI films, magnetic dopants in the bulk can introduce substantial disorder, leading to spatial fluctuations in chemical potential and magnetic exchange gaps [21,45,46]. Consequently, the $C = 1$ QAH effect in uniformly magnetically doped TI films is typically limited to a relatively low critical temperature of ~2 K (refs. [5,12-14,23,25,47-49]). In addition, the Dirac point can lie close to, or overlap with, the bulk valence band, further enhancing dissipative channels [19,50,51]. These inhomogeneities make the effective magnetic exchange gap significantly smaller than the expected magnetic exchange gap inferred from the Curie temperature $T_C$.

To overcome these limitations, modulation doping strategies have been developed, representing a direct implementation of the "topological Legos" approach in which Cr-doped and undoped $(Bi,Sb)_2Te_3$ layers are spatially arranged to control exchange coupling and disorder. In a trilayer heterostructure, thin Cr-doped $(Bi,Sb)_2Te_3$ layers are introduced at the top and bottom surfaces of an undoped $(Bi,Sb)_2Te_3$ spacer layer (Fig. 3a). This design suppresses bulk disorder while maintaining strong exchange coupling at the top and bottom surfaces, thereby reducing spatial fluctuations in chemical potential and the magnetic exchange gaps [22]. This modulation-doping approach represents a key advance in realizing the $C = 1$ QAH state in magnetic TI trilayers, characterized by a single CEC propagating along the sample boundary. At the same time, the bulk and surfaces remain insulating [27,52]. Over the past decade, we have employed MBE to synthesize 3 QL Cr-doped $(Bi,Sb)_2Te_3$/$d$ QL $(Bi,Sb)_2Te_3$/3 QL Cr-doped $(Bi,Sb)_2Te_3$ trilayers (Fig. 3b). Increasing the spacer thickness $d$ decouples the two surfaces while preserving magnetic exchange gaps at each boundary. For a $d = 5$ magnetic TI trilayer, a perfectly quantized zero-magnetic-field Hall resistance $\rho_{yx}(0)$ and a vanishing zero-magnetic-field longitudinal resistance $\rho_{xx}(0)$ were observed at $V_g = 0$ V and $T = 25$ mK, indicating the ferromagnetic order at low temperatures gaps out the top and bottom surface states and the chemical potential lies within the magnetic exchange gaps of both surfaces (Fig. 3c,d) [27]. The $\rho_{yx}$ value remains nearly quantized as $T$ increases to 0.4 K, indicating a robust QAH state with an enhanced critical temperature in modulation-doped magnetic TI heterostructures. These results establish modulation doping as an effective approach for synthesizing high-quality QAH samples.

Increasing the spacer thickness $d$ further reduces the exchange coupling between the top and bottom 3 QL Cr-doped $(Bi,Sb)_2Te_3$ layers, enabling the realization of a 3D QAH insulator. For a $d = 100$ magnetic TI trilayer, a well-quantized $C = 1$ QAH state is observed at the charge neutral point $V_g = V_g^0$ and $T = 25$ mK, with $\rho_{yx}(0) \sim 0.985\ h/e^2$ and $\rho_{xx}(0) \sim 0.003\ h/e^2$ (Fig. 3e,f) [52]. The persistence of the QAH effect in a magnetic TI trilayer with a total thickness of ~106 nm requires that all nonchiral conduction channels remain gapped. This observation further indicates that the side surfaces remain gapped even in the 106-nm-thick magnetic TI trilayer, likely due to the quantum confinement effect associated with the magnetic exchange gaps on the top and bottom surfaces, ensuring that only the chiral edge mode remains conductive. Moreover, the realization of the 3D QAH state in thick magnetic TI trilayers, where the hybridization gap vanishes, is associated with half-quantized surface Hall conductance arising from the axion term in the electromagnetic response, thereby providing a platform for exploring the TME effect and axion physics [53].

### 2.2 High-*C* QAH effect in magnetic TI multilayers

The high-*C* QAH effect has been theoretically proposed in Cr-doped $Bi_2(Se,Te)_3$ films and magnetically doped topological crystalline insulator SnTe [54,55]. In Cr-doped $Bi_2(Se,Te)_3$, the high-*C* QAH effect arises from the inversion of multiple subbands by strong exchange fields [55,56]. In contrast, in SnTe, it is associated with multiple topological surface states [54]. However, experimental realization in these two materials has proven challenging. In Cr-doped $Bi_2(Se,Te)_3$, complex magnetic configurations and residual metallic conduction result in nonsquare $\rho_{xx}$ hysteresis loops [57,58], while in SnTe, the absence of ferromagnetism and the energy misalignment of Dirac cones complicate the realization of high-*C* QAH states [54,59]. Within the "topological Legos" framework (Fig. 2), magnetic TI multilayers provide a more practical and controllable route to high-*C* QAH states. In this approach, multiple $C = 1$ QAH layers are stacked and coupled through intervening normal insulator layers, allowing the total *C* to be engineered through heterostructure design and interlayer coupling [60,61]. Early efforts based on Cr-doped $(Bi,Sb)_2Te_3$/CdSe multilayers were limited by structural incompatibility between wurtzite CdSe and tetradymite $(Bi,Sb)_2Te_3$ (Ref. [62]). This mismatch led to stacking faults and enhanced disorder, resulting in a large $\rho_{xx}$ and a $\rho_{yx}$ that exceeded the quantized value.

As noted above, we realized the well-quantized high-*C* QAH states with *C* from 1 to 5 in [3 QL Cr-doped $(Bi,Sb)_2Te_3$/4 QL $(Bi,Sb)_2Te_3]_m$/3 QL Cr-doped $(Bi,Sb)_2Te_3$ multilayers in 2020 (Fig. 4) [44]. In these magnetic TI multilayers, each 4 QL $(Bi,Sb)_2Te_3$ layer contributes a $C = 1$ QAH channel, while the adjacent 3 QL heavily Cr-doped $(Bi,Sb)_2Te_3$ layers provide both TRS breaking and control of interlayer coupling. As a result, the total *C* scales with the number of 4 QL $(Bi,Sb)_2Te_3$ layers *m*, yielding a quantized $\rho_{yx}(0) = h/(Ce^2)$ with *C* from 1 to 5, accompanied by vanishing $\rho_{xx}(0)$ (Fig. 4). Systematic transport measurements reveal clear quantized Hall plateaus for each $C$, evolving from $h/e^2$ to $h/5e^2$ as additional [3 QL Cr-doped $(Bi,Sb)_2Te_3$/4 QL $(Bi,Sb)_2Te_3$] periods are stacked, demonstrating the parallel contribution of multiple CECs. These results establish magnetic TI multilayers as a scalable and highly controllable platform for engineering high-$C$ QAH states, where both the magnitude and topology of the Hall response can be systematically controlled through layer-by-layer heterostructure engineering.

**2.3 Engineering plateau phase transitions in magnetic TI pentalayers**

Following the realization of the high-$C$ QAH effect in magnetic TI multilayers, it became possible to engineer plateau-to-plateau phase transitions through controlled heterostructure design [44,63,64]. In 3 QL $(Bi,Sb)_{1.73}Cr_{0.27}Te_3$/4 QL $(Bi,Sb)_2Te_3$/3 QL $(Bi,Sb)_{2-x}Cr_xTe_3$/4 QL $(Bi,Sb)_2Te_3$/3 QL $(Bi,Sb)_{1.73}Cr_{0.27}Te_3$ pentalayers, the Cr doping concentration $x$ in the middle magnetic layer can be systematically varied [44,63]. As $x$ increases, a quantum phase transition between the $C = 1$ and $C = 2$ QAH states is realized at zero magnetic field, with $\rho_{yx}$ evolving from $h/e^2$ to $h/2e^2$. This transition is governed by the coupling strength between two adjacent $C = 1$ QAH layers. It is primarily driven by the evolution of the middle magnetic TI layer from a topologically nontrivial to a trivial phase. In the conventional quantum Hall effect, a change in $C$, reflected by the number of CECs, is typically achieved by sweeping $\mu_0H$ or tuning the two-dimensional (2D) carrier density $n_{2D}$ (refs. [65-68]). Since $\mu_0H$ is required for the formation of Landau levels, plateau phase transitions occur only under strong magnetic fields. In contrast, the QAH effect realizes chiral edge transport without an external magnetic field, allowing intrinsic parameters such as magnetic doping to serve as tuning knobs for continuously varying $C$ and accessing plateau phase transitions at $\mu_0H = 0$ T (refs. [12-14,19,20]).

Besides magnetic doping, the thickness of the middle heavily Cr-doped $(Bi,Sb)_2Te_3$ layer provides another effective parameter for engineering plateau phase transitions. In the same pentalayer configuration, varying the thickness $t$ of the middle $(Bi,Sb)_{1.73}Cr_{0.27}Te_3$ layer directly modifies the interlayer coupling between adjacent $C = 1$ QAH layers, enabling a similar transition between the $C = 1$ and $C = 2$ QAH states [44,64]. For $t \leq 1$, the hybridization gap exceeds the magnetic exchange gap for the two middle Dirac surface states, making them topologically trivial and irrelevant to the total $C$. Consequently, only one pair of topologically nontrivial interface states is formed, resulting in a $C = 1$ QAH state. For $t > 1$, the hybridization gap decreases, and beyond a critical thickness, i.e., $t \geq 1.7$, the magnetic exchange gap dominates, resulting in the two inner surface states becoming topologically nontrivial [11,44,58,69,70]. This produces an additional CEC and realizes the $C = 2$ QAH state.

Beyond tuning $C$ in magnetic TI pentalayers, another important direction is the creation of plateau phase transitions with different Chern number change $\Delta C$ in a single magnetic TI multilayer [71,72]. Within the "topological Legos" framework (Fig. 2), this can be achieved by designing asymmetric magnetic TI multilayers in which different magnetic TI layers possess distinct coercive fields, enabling sequential magnetization reversal. In our MBE-grown 3 QL V-doped $(Bi,Sb)_2Te_3$/3 QL $(Bi,Sb)_2Te_3$/2 QL Cr-doped $(Bi,Sb)_2Te_3$/3 QL $(Bi,Sb)_2Te_3$/2 QL Cr-doped $(Bi,Sb)_2Te_3$ pentalayers (Fig. 5a), both $C = 2$ and $C = -1$ QAH states can be stabilized depending on the relative magnetization alignment of the different magnetic TI layers (Fig. 5b) [71]. When the magnetizations of the two 2 QL Cr-doped $(Bi,Sb)_2Te_3$ and the 3 QL V-doped $(Bi,Sb)_2Te_3$ layers are aligned parallel, the sample exhibits a $C = \pm 2$ QAH state with a pair of CECs. In contrast, antiparallel magnetization alignment produces a $C = \pm 1$ QAH state with a single CEC. These magnetic configurations are directly manifested in electrical transport measurements (Fig. 5c). At $\mu_0 H = 0$ T, the $C = 2$ QAH state exhibits a quantized $\rho_{yx} \sim h/2e^2$ together with a nearly vanishing $\rho_{xx}$, indicating the presence of two CECs. Upon sweeping $\mu_0 H$, the sample transitions into the $C =$

-1 QAH state, where $\rho_{yx} \sim -h/e^2$ and $\rho_{xx}$ remains small, consistent with a single CEC. In this scenario, the $C = -1$ QAH effect can be viewed as arising from the coexistence of an axion insulator state and a parallel $C = -1$ QAH state. During the magnetization reversal process, $\rho_{yx}$ evolves between these quantized values, while $\rho_{xx}$ exhibits pronounced peaks at the transition fields corresponding to the switching of different magnetic TI layers. As a result, plateau phase transitions with $\Delta C = 1$ and $\Delta C = 3$ can be realized within the same device. These transitions are further visualized in conductivity flow diagrams, where the evolution of ($\sigma_{xy}$, $\sigma_{xx}$) traces semicircular trajectories, confirming two types of plateau phase transitions with $\Delta C = 1$ and $\Delta C = 3$ (Refs. [13,73]). The radius of each semicircle curve is determined by $\Delta C$, directly reflecting the underlying plateau phase transition in QAH insulators.

### 2.4 Mesoscopic QAH devices and electrical control of QAH states

Over the past decade, an important direction in the study of magnetically doped QAH materials has been the development of mesoscopic devices and the electrical control of QAH states in nanoscale structures. Before 2022, miniaturized QAH devices were typically patterned using either mechanical scratching [12,13,19,20,23,27,28,48,52,63,71] or photolithography [24,29,47,49,74-77], with the minimum Hall bar width $w$ limited to ~10 μm (ref. [24]). In these QAH devices, the two CECs on opposite sides of the sample remain well separated, so the QAH effect usually exhibits quantized $\rho_{yx}$ and vanishing $\rho_{xx}$. However, for potential applications, it is essential to miniaturize QAH devices to dimensions within the quantum coherence length. Moreover, when the width $w$ of the QAH Hall bar becomes smaller than $2d_0$, where $d_0$ is the CEC width, the two CECs are predicted to couple with each other, leading to the breakdown of the QAH effect [78]. By combining photolithography and electron-beam lithography, the QAH effect is observed in magnetic TI Hall bar devices with $w \approx 600$ nm (ref. [79]) and $w \approx 160$ nm (ref. [80]). We further reduce the minimum

Hall bar device width $w$ to ≈72 nm using pure electron-beam lithography on 3 QL Cr-doped $(Bi,Sb)_2Te_3$/4 QL $(Bi,Sb)_2Te_3$/3 QL Cr-doped $(Bi,Sb)_2Te_3$ trilayers (Fig. 6a) [81]. Remarkably, the QAH state persists even at $w$ ~72 nm, with $\rho_{yx}(0)$ ~0.9517 $h/e^2$ and $\rho_{xx}(0)$ ~0.1911 $h/e^2$ (Fig. 6b). The ratio $\rho_{yx}(0)/\rho_{xx}(0)$ corresponds to an anomalous Hall (AH) angle α ~78.65°, indicating the chiral edge transport still dominates over bulk transport. The finite $\rho_{xx}(0)$ likely originates from confinement-induced coupling between the two opposite CECs.

In QAH insulators, the locking between magnetization and edge current chirality provides an opportunity to electrically switch the edge current chirality by reversing the magnetization direction through spin-orbit torque (SOT) (refs. [10,82]). Realizing SOT switching in QAH insulators requires two key conditions: (1) a sufficiently large SOT effect in the QAH structure and (2) a sufficiently high current density to switch the magnetization. The first condition can be achieved by engineering an asymmetric chemical potential between the top and bottom surfaces of a QAH trilayer through their different local environments, while the second can be achieved by fabricating narrow QAH Hall bar devices that enhance the current density [81]. Since the top and bottom magnetic TI layers in QAH trilayers possess similar magnetic moments, their magnetizations can switch simultaneously under SOT (Fig. 6c), producing a QAH insulator with opposite edge-current chirality after SOT switching.

As noted above, using 3 QL Cr-doped $(Bi,Sb)_2Te_3$/4 QL $(Bi,Sb)_2Te_3$/3 QL Cr-doped $(Bi,Sb)_2Te_3$ trilayers, we employed electron-beam lithography to fabricate QAH Hall bar devices with $w$ down to ≈72 nm (ref. [83]). We systematically investigated the electrical control of the QAH state in Hall bar devices with different $w$ and found that the electrical switching of the QAH state can be realized only within the range 1 μm ≤ $w$ ≤ 10 μm (ref. [83]). For Hall bar devices with $w$ > 10 μm, the current density is insufficient for magnetization switching. For devices with $w$ < 1 μm, the substantial contact resistance significantly reduces the efficiency of SOT-induced magnetization switching. Figure 6d shows SOT switching between $C = \pm1$ QAH states in a $w = 5$ μm Hall bar device. When a current pulse $I_{\mathrm{pulse}} \sim -200$ μA is applied at $V_g = V_g^0$ under an in-plane magnetic field $\mu_0 H_{\parallel} = +0.05$ T, the $C = 1$ QAH state switches to the $C = -1$ QAH state. Before SOT switching, the $C = 1$ QAH state exhibits $\rho_{yx}(0) \sim 1.0005 \pm 0.0020\ h/e^2$ and $\rho_{xx}(0) \sim 0.0220 \pm 0.0007\ h/e^2$ at $V_g = V_g^0$, corresponding to a right-handed edge current. After SOT switching, the device exhibits $\rho_{yx}(0) \sim -1.0008 \pm 0.0011\ h/e^2$ and $\rho_{xx}(0) \sim 0.0562 \pm 0.0006\ h/e^2$, corresponding to a left-handed edge current in the $C = -1$ QAH state (Fig. 6d). Starting from the initial $C = 1$ QAH state, four successive SOT switching events of the edge current chirality are achieved in the $w = 5$ μm Hall bar device by reversing either $\mu_0 H_{\parallel}$ or $I_{\mathrm{pulse}}$ (Fig. 6e). These results establish that the chirality of the QAH edge current can be controllably switched through current-pulse injection under an in-plane magnetic field via the SOT effect.

**3. Axion insulator state**

The realization of an axion insulator state in a TI requires three conditions [6,84,85]: (1) the TI film must be in the 3D regime; (2) all surfaces must be gapped, with the chemical potential lying within the magnetic exchange gaps; and (3) the TI bulk must preserve TRS or inversion symmetry so that $\theta = \pi$ is maintained. A recent study has interpreted the observation of a zero Hall conductance plateau (ZHCP) as evidence for an axion insulator arising from antiparallel magnetization alignment of the top and bottom Cr-doped $(\mathrm{Bi,Sb})_2\mathrm{Te}_3$ layers [86]. However, subsequent magnetic domain imaging measurements failed to find evidence of antiparallel magnetization alignment at any external $\mu_0 H$ (ref. [87]). Therefore, an alternative approach based on asymmetric magnetic TI trilayers (Figs. 2c and 7a, b) has been developed to realize axion insulator states [6,84,85]. In these magnetic TI trilayers, the top and bottom $(\mathrm{Bi,Sb})_2\mathrm{Te}_3$ layers are doped with two different magnetic ions, specifically Cr and V. If the interlayer exchange coupling is substantially smaller than the difference between the coercive fields of the two magnetic TI layers, an antiparallel magnetization alignment can emerge when $\mu_0 H$ lies between the two coercive fields, thereby supporting an axion insulator state (Fig. 7a). Such behavior has been demonstrated in our MBE-grown 3 QL V-doped $(Bi,Sb)_2Te_3$/$d$ QL $(Bi,Sb)_2Te_3$/3 QL Cr-doped $(Bi,Sb)_2Te_3$ trilayers with $d \geqslant 3$ (Fig. 7b) [16,88,89].

For the $d = 4$ magnetic TI trilayer, $\sigma_{xx}$ exhibits sharp peaks at the coercive fields, corresponding to the sequential magnetization reversal of the Cr-doped and V-doped $(Bi,Sb)_2Te_3$ layers (Fig. 7d). The $\mu_0 H$ dependence of $\sigma_{xy}$ shows a two-step transition from $e^2/h$ to 0 and then to $-e^2/h$, with a well-defined ZHCP appearing between the two coercive fields (Fig. 7c). The simultaneous observation of a ZHCP and suppressed $\sigma_{xx}$ when $\mu_0 H$ lies between two coercive fields results from the cancellation of the Hall conductance contributions from the top and bottom surfaces under antiparallel magnetization alignment, consistent with an axion insulator state (Figs. 2c and 7a, b). This interpretation is further supported by magnetic force microscopy measurements, which directly reveal a two-step transition similar to the ZHCP and confirm the independent magnetization reversal of the Cr-doped and V-doped $(Bi,Sb)_2Te_3$ layers. Unlike the ZHCP induced by surface state hybridization observed in uniformly doped $(Bi,Sb)_2Te_3$ films [30,31,90], the mechanism here relies on independently controlled surface magnetizations, providing a clearer and more controllable platform for exploring the TME effect and axion physics.

Even in a magnetic TI trilayer, a ZHCP can also arise from surface state hybridization when the middle undoped $(Bi,Sb)_2Te_3$ layer is in the 2D regime [103,106,107]. Therefore, realizing an axion insulator state in a sufficiently thick asymmetric magnetic TI trilayer is essential. To this end, we used MBE to synthesize a $d = 100$ asymmetric magnetic TI trilayer. The total sample thickness is

~106 nm, which prevents the formation of a hybridization gap between the two surface states of the middle $(Bi,Sb)_2Te_3$ layer, since the surface state penetration depth in $(Bi,Sb)_2Te_3$ is only a few nanometers [30,31]. Electrical transport measurements reveal a robust ZHCP between the distinct coercive fields of the top 3 QL V-doped $(Bi,Sb)_2Te_3$ layer and the bottom 3 QL Cr-doped $(Bi,Sb)_2Te_3$ layer, accompanied by nearly vanishing $\sigma_{xx}$ (Fig. 7e,f). The ZHCP and the corresponding nearly vanishing $\sigma_{xx}$ result from the cancellation of the top and bottom surface contributions under antiparallel magnetization alignment, further confirming the realization of the axion insulator state in the $d = 100$ magnetic TI trilayer.

**4. Parity anomaly state and half-quantized chiral edge transport**

In contrast to integer quantization realized in QAH insulators, the $C = 1/2$ parity anomaly state describes a quantum field theoretical phenomenon in which a single massless Dirac fermion coupled to an electromagnetic $U(1)$ gauge field breaks parity symmetry upon quantization and generates a half-integer $\sigma_{xy}$ when an infinitesimal mass term is introduced [91-98]. 3D TIs provide a natural platform for exploring parity anomaly physics because each TI surface hosts a single gapless Dirac fermion [15,99,100]. When magnetic exchange interactions open a gap in one surface state while leaving the opposite surface gapless, a half-quantized $\sigma_{xy}$ is predicted [101]. Building on this insight, experimental evidence for a $C = 1/2$ parity anomaly state has been reported in magnetic TI bilayers [102], where magnetic doping near the top surface gaps the top Dirac surface state while the bottom surface remains gapless (Fig. 8a). However, in these magnetic TI bilayers, the half-quantized response is sensitive to external $\mu_0H$. The existence of corresponding half-quantized chiral edge transport remains unclear.

To overcome this limitation, the asymmetric magnetic TI trilayers discussed above provide a particularly effective platform for realizing the $C = 1/2$ parity anomaly state [103-106]. In our MBE-grown 3 QL V-doped $(Bi,Sb)_2Te_3$/6 QL $(Bi,Sb)_2Te_3$/3 QL Cr-doped $(Bi,Sb)_2Te_3$ trilayers, the middle 6 QL undoped $(Bi,Sb)_2Te_3$ spacer layer weakens the interlayer exchange coupling while preserving distinct magnetic anisotropies between the top 3 QL V-doped $(Bi,Sb)_2Te_3$ layer and the bottom 3 QL Cr-doped $(Bi,Sb)_2Te_3$ layer (Fig. 8b) [16,17]. As a result, the magnetizations of the two magnetic TI layers can be tuned independently, enabling selective control of the Dirac surface states. At $\mu_0 H_x$ = 0 T, the magnetizations of the top 3 QL V-doped $(Bi,Sb)_2Te_3$ layer and the bottom 3 QL Cr-doped $(Bi,Sb)_2Te_3$ layer are aligned in parallel, opening magnetic exchange gaps on both the top and bottom surfaces of the middle 6 QL $(Bi,Sb)_2Te_3$ layer. In this regime, the two gapped surfaces each contribute a half-quantized $\sigma_{xy}$, giving rise to a $C = 1$ QAH state (Fig. 8c). As $\mu_0 H_x$ increases, the difference in anisotropy fields and the weak interlayer exchange coupling between the top 3 QL V-doped $(Bi,Sb)_2Te_3$ layer and the bottom 3 QL Cr-doped $(Bi,Sb)_2Te_3$ layers allow the magnetization of the bottom 3 QL Cr-doped $(Bi,Sb)_2Te_3$ layer to tilt into the plane, while the top 3 QL V-doped $(Bi,Sb)_2Te_3$ layer maintains an out-of-plane magnetization. This produces a configuration in which the bottom surface state of the middle 6 QL $(Bi,Sb)_2Te_3$ layer becomes gapless, whereas the top surface state remains gapped [11,12,107,108]. In this regime, a half-quantized Hall conductance $\sigma_{xy} \sim e^2/2h$ is observed, identifying the $C = 1/2$ parity anomaly state. At the same time, $\sigma_{xx}$ remains finite because of dissipative transport associated with the gapless surface (Fig. 8c).

Upon further increasing $\mu_0 H_x$, the magnetization of the top 3 QL V-doped $(Bi,Sb)_2Te_3$ layer is also driven into the plane, rendering both surface states gapless and producing a $C = 0$ state characterized by vanishing $\sigma_{xy}$ and enhanced $\sigma_{xx}$ (Fig. 8c). The evolution of the transport properties as a function of $\mu_0 H_x$ therefore reveals three distinct regimes: $\sigma_{xy}$ decreases from $e^2/h$ to $e^2/2h$ and eventually to 0, while $\sigma_{xx}$ exhibits corresponding changes associated with the progressive closing of the surface magnetic exchange gaps. This sequence reflects a transition from a $C = 1$ QAH state to a $C = 1/2$ parity anomaly state, then to a $C = 0$ metallic state. The realization of the $C = 1/2$ parity anomaly state and a robust half-quantized $\sigma_{xy}$ plateau in MBE-grown asymmetric magnetic TI trilayers provides new insight into Berry-phase phenomena associated with a single massive Dirac fermion and opens a route toward further exploration of parity-anomaly-related topological transport phenomena. Furthermore, the observation of nonlocal and nonreciprocal transport in the $C = 1/2$ parity anomaly state provides compelling evidence for the existence of a half-quantized chiral edge transport. This half-quantized $\sigma_{xy}$ plateau and the half-quantized chiral edge transport arise from bulk-boundary correspondence and remain robust against parameter variations [109-111].

## 5. Discussion and outlook

As discussed above, magnetic TI multilayers have emerged as a versatile platform for realizing designer quantum states, such as $C = 1$ and high-$C$ QAH insulators, axion insulators, and $C = ½$ parity anomaly states. Looking forward, the "topological Legos" approach (Fig. 2) provides a powerful route toward engineering new phases beyond those realized to date. By further tuning layer thickness, magnetic doping, interlayer coupling, dimensionality, and chemical potential, magnetic TI multilayers may provide access to several unexplored regimes of topological matter. Below, we highlight three promising directions for future research.

### 5.1 QAH metal phase

As noted above, the QAH effect is traditionally understood as an insulating state characterized by quantized $\sigma_{xy}$ and vanishing $\sigma_{xx}$ [11-15]. However, recent theoretical studies [112-115] have predicted that the QAH phase can persist even in metallic systems where a full bulk gap is absent. In this scenario, a ferromagnetic metal may host coexisting CECs and isotropic bulk conduction without a global energy gap, giving rise to quantized $\sigma_{xy}$ despite finite $\sigma_{xx}$ [112-115]. TI/magnetic TI/TI trilayers have been predicted to host a metallic QAH state in which $\sigma_{xy}$ remains quantized while $\sigma_{xx}$ is finite, and conventional CECs are not strictly required [112,113]. Furthermore, dephasing effects may play a crucial role in stabilizing this regime by suppressing ballistic metallic transport and driving the system into a diffusive regime, where a well-defined local electric field allows $\sigma_{xy}$ to approach the quantized value even in the presence of finite $\sigma_{xx}$. By blurring the conventional boundary between quantized Hall insulators and trivial metals, the QAH metal phase provides a broader framework for understanding topological transport in gapless and potentially disordered magnetic systems [112]. Future studies should focus on identifying clear transport signatures that distinguish a QAH metal from a dissipative QAH insulator, including scaling behavior, nonlocal transport, mesoscopic edge-channel imaging, and finite-frequency responses. Establishing the QAH metal phase would bridge the gap between isolated QAH insulators and more complex topological metallic states, providing a new route towards tunable chiral transport at $\mu_0 H = 0$ T.

### 5.2 Synthetic magnetic Weyl semimetal phase

By extending the multilayer design approach, alternating magnetic and nonmagnetic TI layers can be viewed as an artificial crystal grown along the out-of-plane direction, in which interlayer coupling plays the role of momentum-dependent dispersion. By tuning the TI spacer thickness, magnetic exchange gap, and coupling between adjacent TI layers, the system can, in principle, be driven from a stack of 2D QAH insulators into a 3D topological phase hosting Weyl nodes [60,61,116]. Compared with naturally occurring Weyl semimetals, synthetic magnetic Weyl semimetals based on magnetic TI multilayers offer a much higher degree of tunability, because their band topology, magnetic configuration, and dimensional crossover can be engineered layer by layer. Experimentally, key challenges include improving crystalline quality over many repeated layers, suppressing residual bulk conduction, and directly resolving Weyl physics through transport, optical, and spectroscopic probes. Future work should aim to identify hallmark signatures of the magnetic Weyl semimetal phase, such as AH responses controlled by Weyl-node separation, chiral-anomaly-related transport, surface Fermi-arc states, and thickness-dependent dimensional crossover. The realization of synthetic magnetic Weyl semimetals would extend magnetic TI multilayers beyond 2D QAH insulators, establishing them as a versatile platform for engineering designer 3D topological quantum phenomena.

### 5. 3 TME effect in thick magnetic TI trilayers

Thick magnetic TI trilayers provide particularly promising platforms for exploring the TME effect because they can enter a true 3D regime while maintaining gapped surfaces and an insulating bulk. In a thick asymmetric magnetic TI trilayer exhibiting the axion insulator state (Fig. 7e,f), antiparallel magnetization alignment between the top and bottom magnetic layers is expected to produce canceling half-quantized surface Hall conductances and a quantized TME response [6,84,85,117,118]. In contrast, in a thick symmetric magnetic TI trilayer exhibiting the 3D QAH effect (Fig. 3e,f), parallel magnetization alignment leads to a CEC associated with the sum of the half-quantized surface responses. Demonstrating the TME effect in these two types of magnetic TI trilayers will require going beyond dc transport signatures, such as the ZHCP or quantized Hall resistance, towards direct electromagnetic probes of the quantized response. Future experiments may include optical Faraday and Kerr rotation, scanning magnetic probes, and device geometries designed to detect electric-field-induced magnetization or magnetic-field-induced polarization [6,119-121]. Success in this direction would establish the magnetic TI trilayer as a powerful platform for probing fundamental TME phenomena.

**CRediT authorship contribution statement:**

**Deyi Zhuo**: Writing – original draft, Investigation, Visualization, Validation, Resources,

Conceptualization.

**Han Tay**: Writing – original draft, Investigation, Visualization, Validation, Resources, Conceptualization.

**Cui-Zu Chang**: Writing – review & editing, Investigation, Validation, Supervision, Project administration, Funding acquisition, Conceptualization.

**Declaration of competing interest:** The authors declare that they have no known competing financial interests or personal relationships that could have influenced the work reported in this review article.

**Acknowledgments:** The authors thank D. Xiao, J. Jiang, Y. -F. Zhao, R. Zhang, L. -J. Zhou, Z. -J. Yan, B. Zhang, X. Liu, F. Wang, M. Kayyalha, H. Yi, B. Chen, W. Wu, V. Law, C.-X. Liu, C.-Z. Chen, M. H. W. Chan, N. Samarth, and X. Xu for long-term collaborations on the topics covered in this review article. C.-Z. C. acknowledges support from the DOE grant (DE-SC0023113), ARO grant (W911NF2210159), NSF award (DMR-2241327), Penn State NSF-MRSEC (DMR-2011839), and ONR award (N000142412133), and the Gordon and Betty Moore Foundation's EPiQS Initiative (GBMF9063 to C.-Z. C).

**Figures and figure captions:**

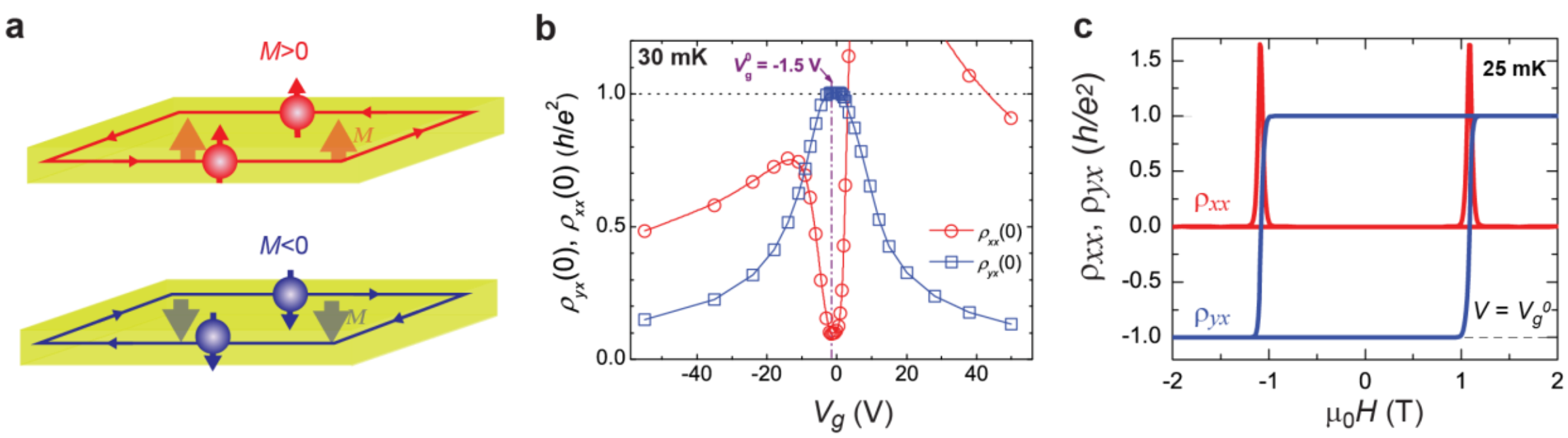

**Fig. 1| *C* = 1 QAH effect in uniformly Cr- and V-doped $(Bi,Sb)_2Te_3$ films. a,** Schematic of a QAH insulator with right (left) handed chiral edge current for positive (negative) magnetization. **b,** $V_g$-dependent $\rho_{yx}(0)$ (blue square) and $\rho_{xx}(0)$ (red circle) of a 5 QL Cr-doped $(Bi,Sb)_2Te_3$ film measured at $T = 30$ mK, from ref. [12]. **c,** $\mu_0H$-dependent $\rho_{yx}$ (blue) and $\rho_{xx}$ (red) of a 4 QL V-doped $(Bi,Sb)_2Te_3$ film measured at $V_g = V_g^0$ and $T = 25$ mK, from ref. [19].

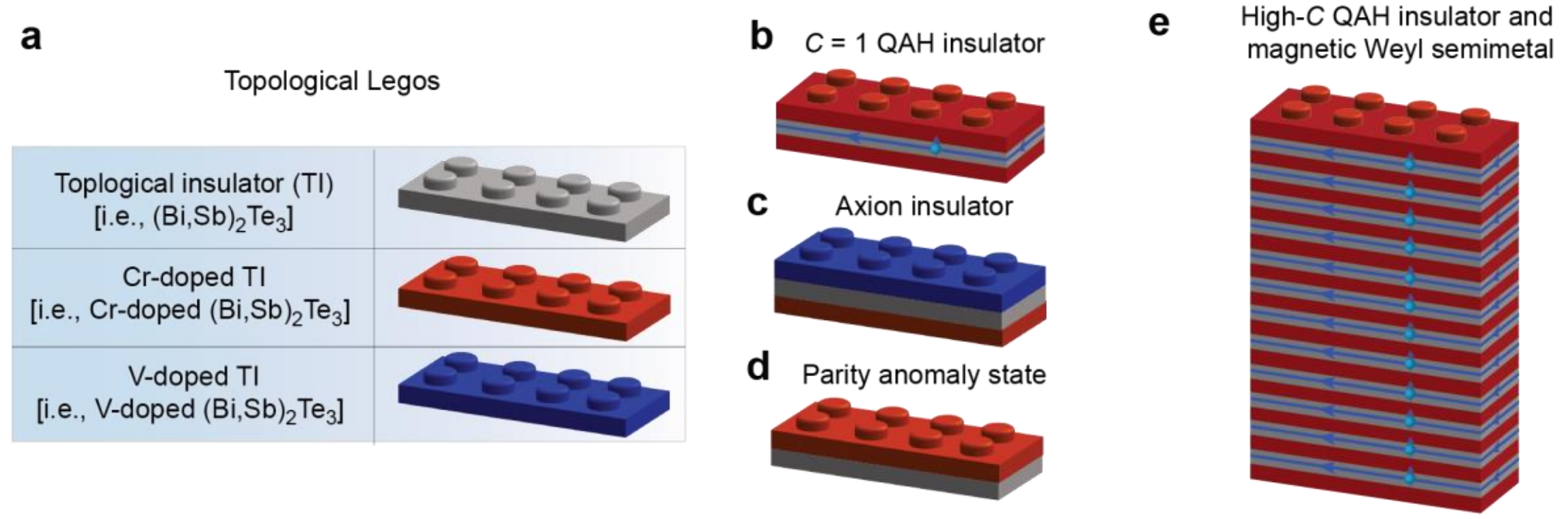

**Fig. 2| Topological Legos. a**, Individual topological building blocks used to design quantum phases, including $C$ = 1 QAH insulator, axion insulator, parity anomaly state, high-$C$ QAH insulator state, and magnetic Weyl semimetal phase. **b, c,** $C$ = 1 QAH insulator (**b**) and axion insulator (**c**) states in magnetic TI trilayers. **d,** Parity anomaly state in a magnetic TI bilayer. **e,** High-$C$ QAH insulator state and magnetic Weyl semimetal phase in magnetic TI superlattices.

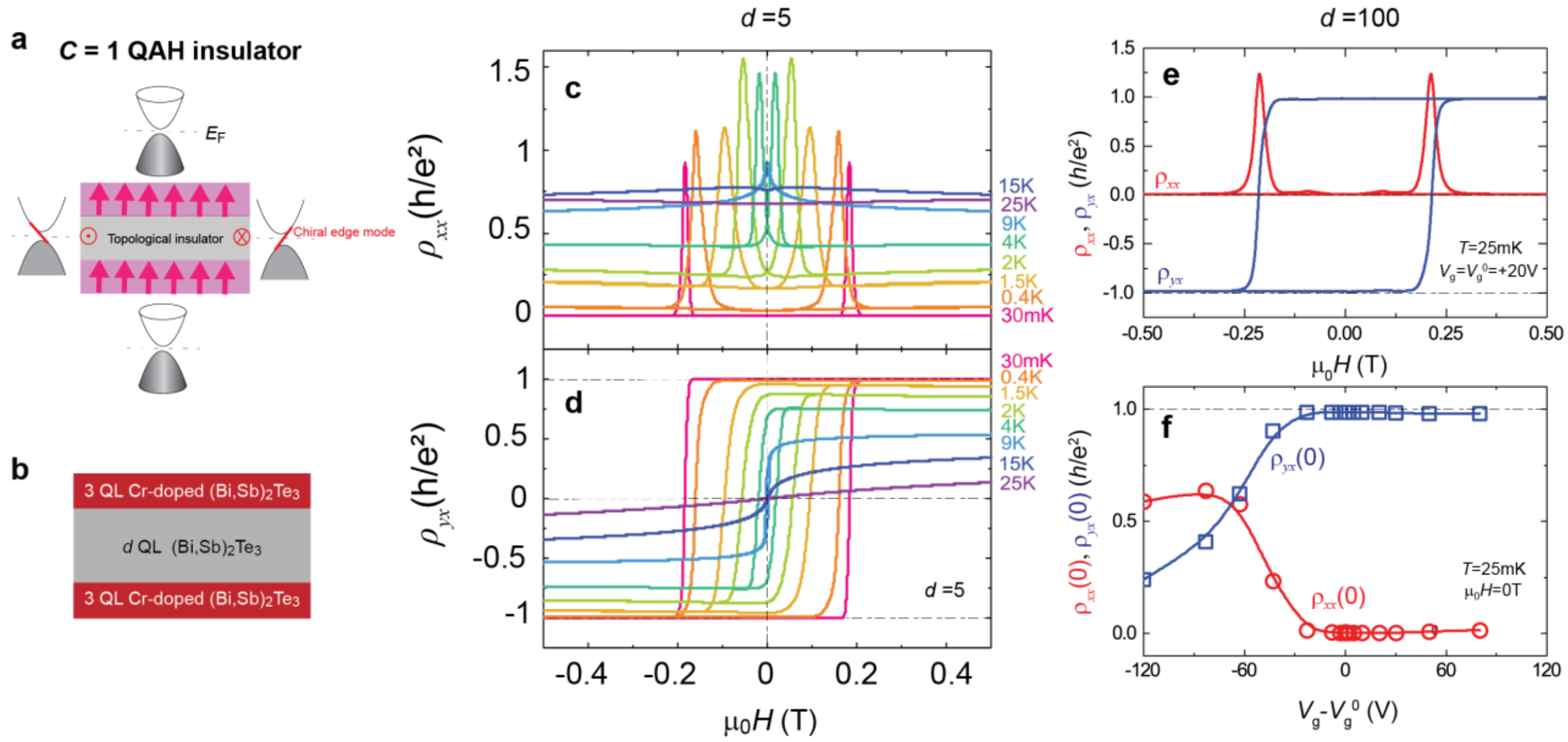

**Fig. 3| $C = 1$ QAH effect in magnetic TI trilayers. a,** Schematic of a $C = 1$ QAH trilayer with a chiral edge current propagating along the sample boundary. **b**, Side view of a symmetric magnetic TI trilayer. **c, d,** $\mu_0H$-dependent $\rho_{xx}$ (**c**) and $\rho_{yx}$ (**d**) of a $d = 5$ magnetic TI trilayer measured at $V_g = 0$ V and $T = 30$ mK, from ref. [27]. **e,** $\mu_0H$-dependent $\rho_{xx}$ (red) and $\rho_{yx}$ (blue) of a $d = 100$ magnetic TI trilayer measured at $V_g = V_g^0$ and $T = 25$ mK, from ref. [52]. **f,** ($V_g$-$V_g^0$)-dependent $\rho_{yx}(0)$ (blue square) and $\rho_{xx}(0)$ (red circle) of the $d = 100$ magnetic TI trilayer measured at $\mu_0H = 0$ T and $T = 25$ mK, from ref. [52].

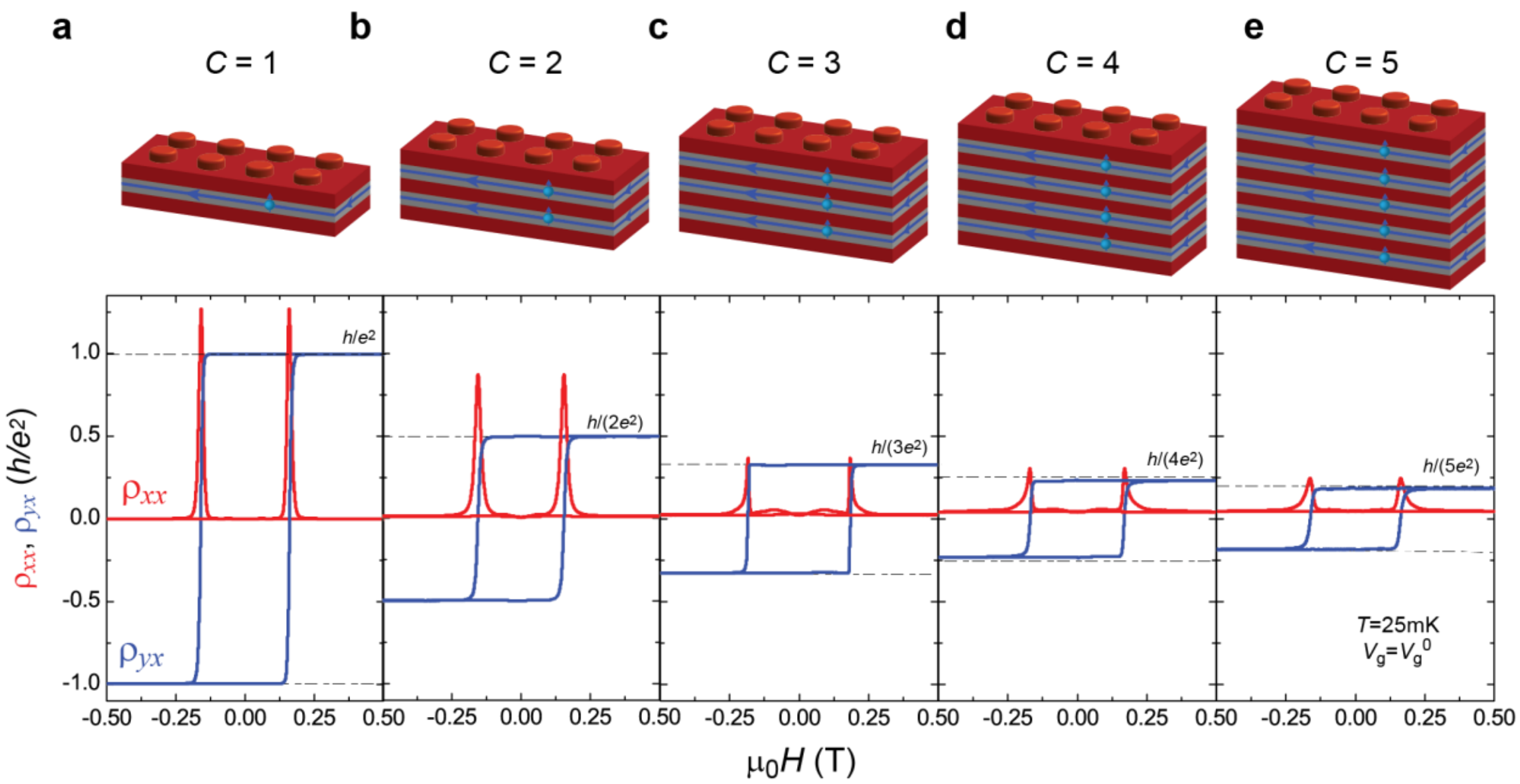

**Fig. 4| High-$C$ QAH effect in magnetic TI multilayers. a-e,** Top panels: Lego-style schematics of magnetic TI multilayer structures hosting QAH states with $C = 1$ to $C = 5$. The red Lego block represents 3 QL heavily Cr-doped $(Bi,Sb)_2Te_3$, and the gray Lego block represents 4 QL $(Bi,Sb)_2Te_3$. Bottom panels: $\mu_0H$-dependent $\rho_{xx}$ (red) and $\rho_{yx}$ (blue) in the corresponding magnetic TI multilayers measured at $V_g = V_g^0$ and $T = 25$ mK, from ref. [44].

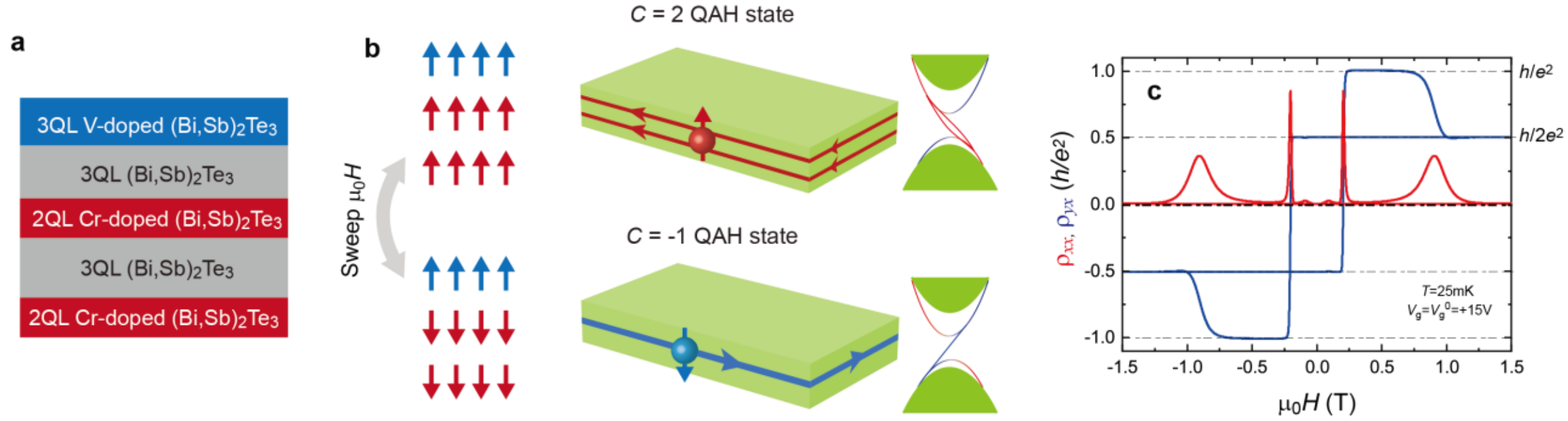

**Fig. 5| Engineering plateau phase transitions in magnetic TI multilayers. a,** Side view of an asymmetric magnetic TI pentalayer. **b,** Schematic of the quantum phase transition between $C = 2$ and $C = -1$ QAH states. **c,** $\mu_0 H$-dependent $\rho_{xx}$ (red) and $\rho_{yx}$ (blue) at $V_g = V_g^0 = +15$ V and $T = 25$ mK, from ref. [71].

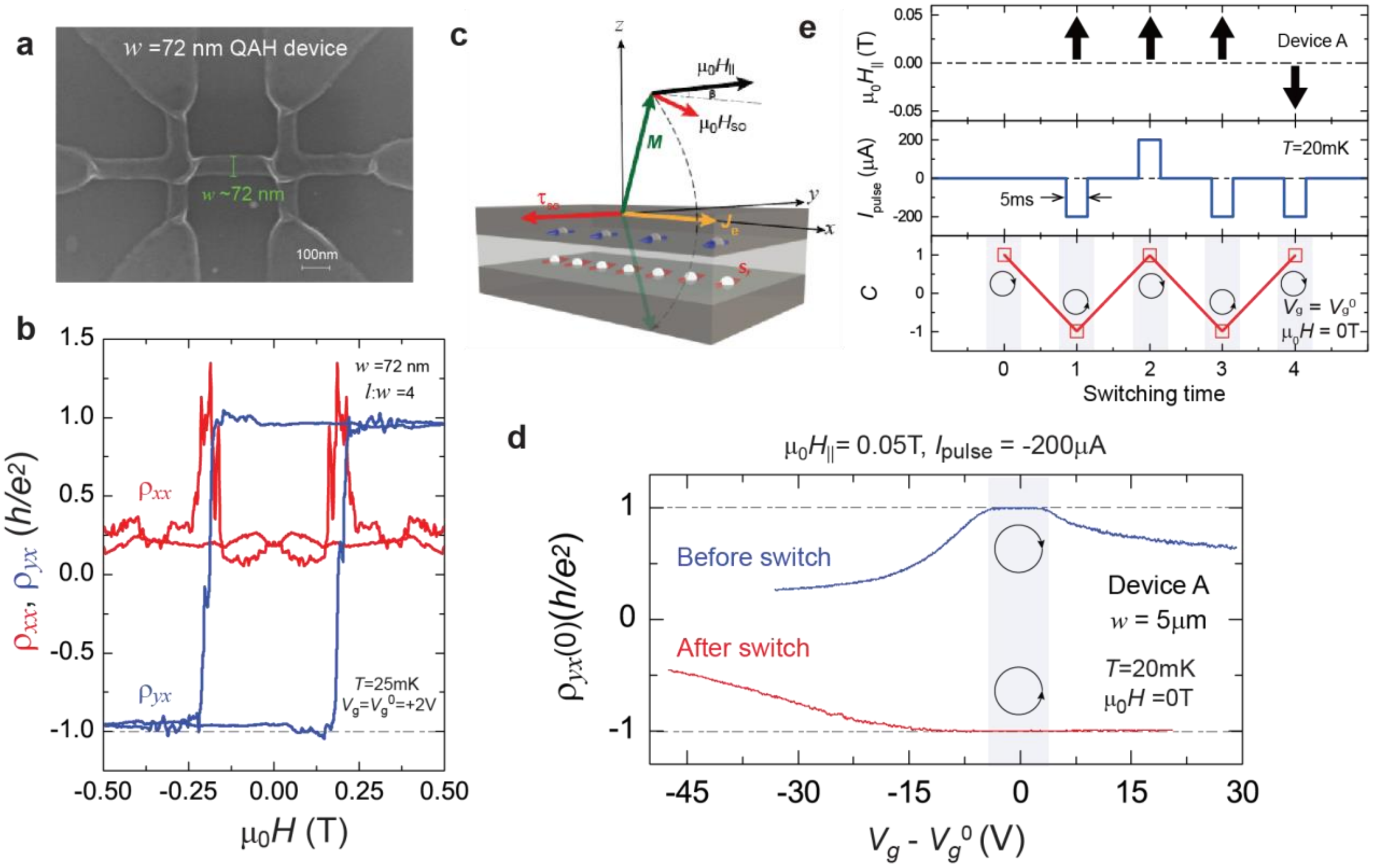

**Fig. 6| Current-pulse-induced switching of QAH states in magnetic TI trilayers. a,** Scanning electron microscope image of a 3 QL Cr-doped $(Bi,Sb)_2Te_3$/4 QL $(Bi,Sb)_2Te_3$/3 QL Cr-doped $(Bi,Sb)_2Te_3$ trilayer with a width of ≈ 72 nm. **b,** $\mu_0 H$-dependent $\rho_{xx}$ (red) and $\rho_{yx}$ (blue) of the $w \sim$ 72 nm device measured at $V_g = V_g^0$ and $T$ = 25 mK, from ref. [81]. **c,** Schematic of SOT-induced magnetization switching in a magnetic TI trilayer under an in-plane magnetic field. **d,** ($V_g$-$V_g^0$)-dependent $\rho_{yx}(0)$ measured before (blue) and after (red) magnetization switching in a 3 QL Cr-doped $(Bi,Sb)_2Te_3$/4 QL $(Bi,Sb)_2Te_3$/3 QL Cr-doped $(Bi,Sb)_2Te_3$ heterostructure with $w$ ≈ 5 μm, from ref. [83]. **e,** Continuous switching of QAH states measured at $T$ = 20 mK. The chirality of the edge current can be reversed by changing the direction of either the in-plane magnetic field or the current pulse, from ref. [83].

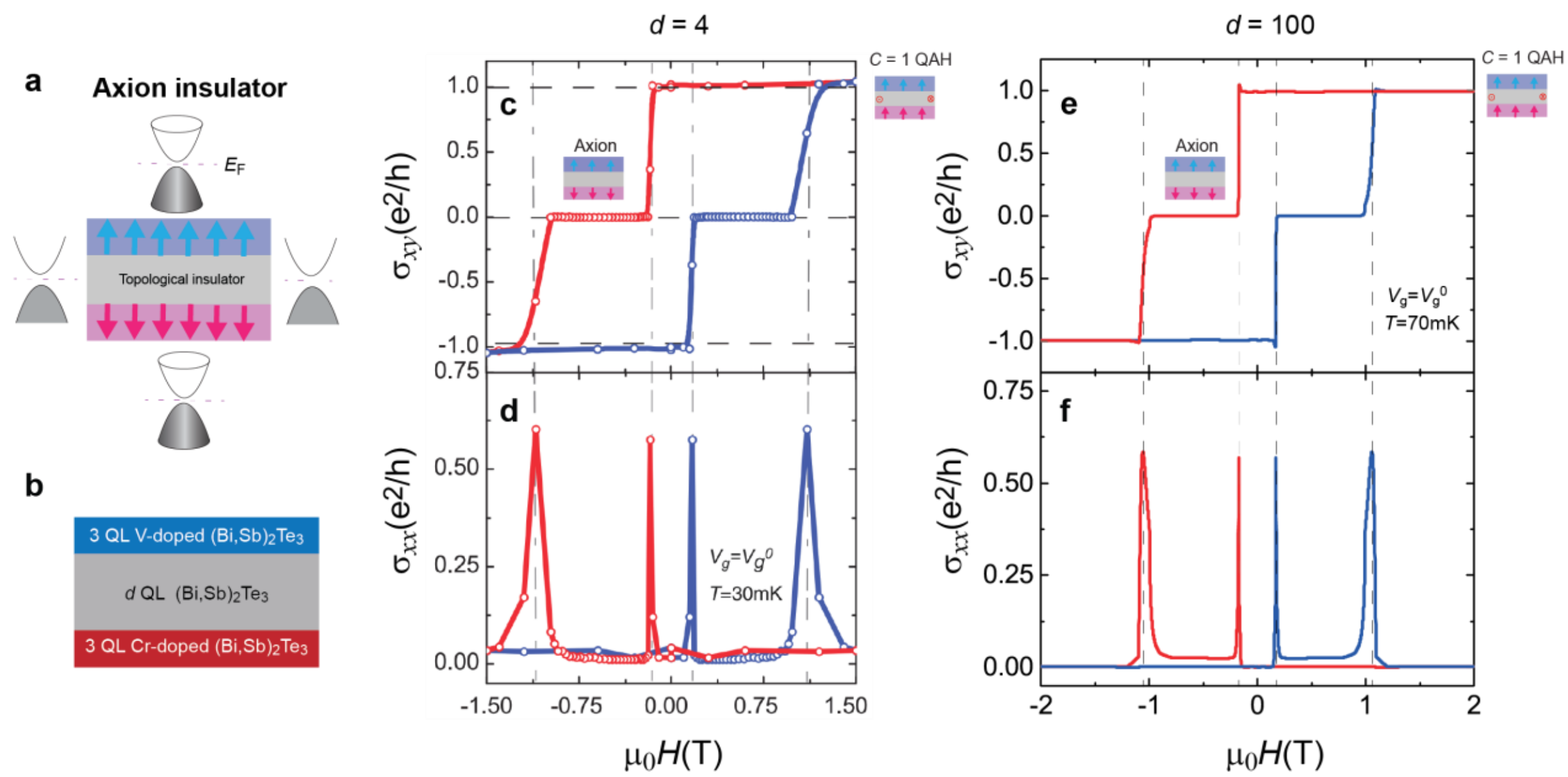

**Fig. 7| Axion insulator state in asymmetric magnetic TI trilayers. a,** Schematic of an asymmetric magnetic TI trilayer with antiparallel magnetization alignment. **b**, Side view of an asymmetric magnetic TI trilayer. **c, d,** $\mu_0 H$-dependent $\sigma_{xy}$ (**c**) and $\sigma_{xx}$ (**d**) of a $d$ = 4 asymmetric magnetic TI trilayer measured at $V_g = V_g^0$ and $T$ = 30 mK, from ref. [16]. **e, f,** $\mu_0 H$-dependent $\sigma_{xy}$ (**e**) and $\sigma_{xx}$ (**f**) of a $d$ = 100 asymmetric magnetic TI trilayer measured at $V_g = V_g^0$ and $T$ = 70 mK, from ref. [88].

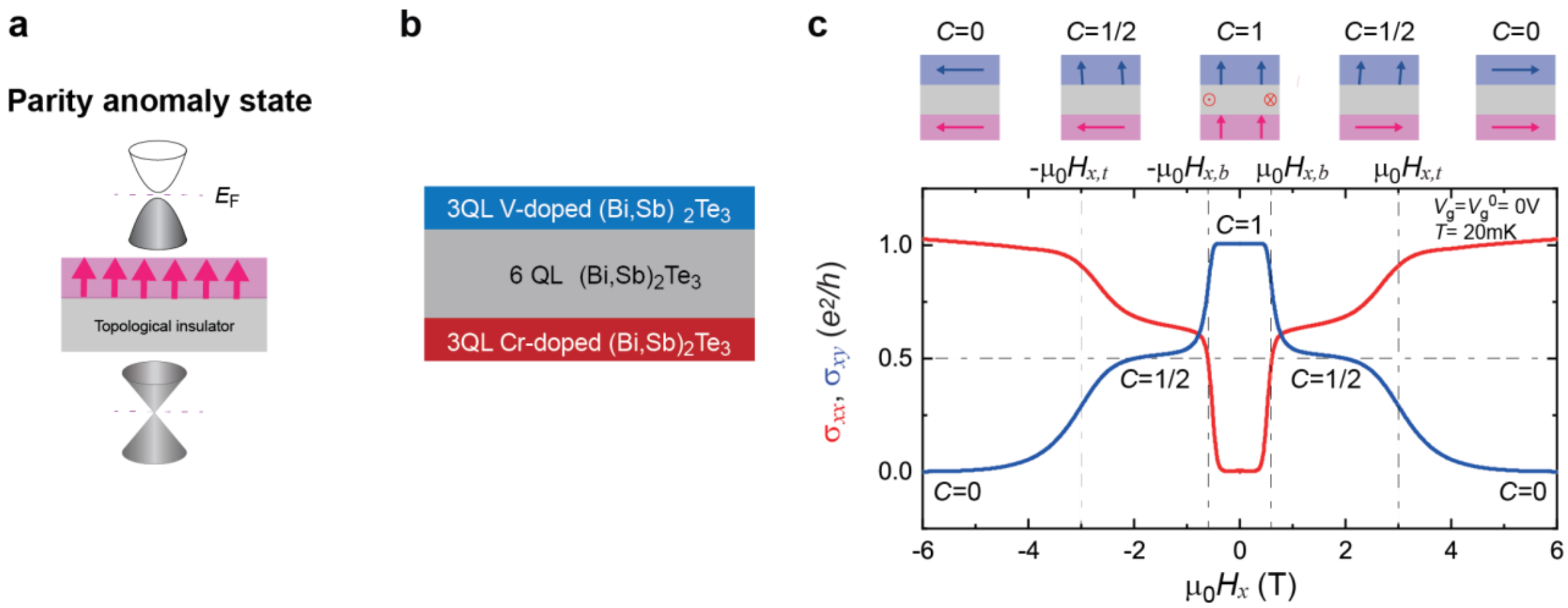

**Fig. 8| Parity anomaly state in magnetic TI bilayers and trilayers. a,** Schematic of a magnetic TI bilayer. **b,** Side view of an asymmetric magnetic TI trilayer. **c,** $\mu_0 H_x$-dependent $\sigma_{xx}$ (red) and $\sigma_{xy}$ (blue) measured at $V_g = V_g^0 = 0$ V and $T = 20$ mK, from ref. [103].

## References

1 Hasan, M. Z. & Kane, C. L. Colloquium: Topological insulators. *Rev. Mod. Phys.* **82**, 3045-3067 (2010).

2 Qi, X.-L. & Zhang, S.-C. Topological insulators and superconductors. *Rev. Mod. Phys.* **83**, 1057-1110 (2011).

3 Moore, J. E. The birth of topological insulators. *Nature* **464**, 194-198 (2010).

4 Chang, C.-Z., Liu, C.-X. & MacDonald, A. H. Colloquium: Quantum anomalous Hall effect. *Rev. Mod. Phys.* **95**, 011002 (2023).

5 Liu, C.-X., Zhang, S.-C. & Qi, X.-L. The quantum anomalous Hall effect: theory and experiment. *Annu. Rev. Condens. Matter Phys.* **7**, 301-321 (2016).

6 Qi, X.-L., Hughes, T. L. & Zhang, S.-C. Topological field theory of time-reversal invariant insulators. *Phys. Rev. B* **78**, 195424 (2008).

7 Chang, C.-Z. & Li, M. Quantum anomalous Hall effect in time-reversal-symmetry breaking topological insulators. *J. Phys. Condens. Matter* **28**, 123002 (2016).

8 Weng, H., Yu, R., Hu, X., Dai, X. & Fang, Z. Quantum anomalous Hall effect and related topological electronic states. *Adv. Phys.* **64**, 227-282 (2015).

9 He, K., Wang, Y. & Xue, Q.-K. Topological materials: quantum anomalous Hall system. *Annu. Rev. Condens. Matter Phys.* **9**, 329-344 (2018).

10 Tokura, Y., Yasuda, K. & Tsukazaki, A. Magnetic topological insulators. *Nat. Rev. Phys.* **1**, 126-143 (2019).

11 Yu, R., Zhang, W., Zhang, H.-J., Zhang, S.-C., Dai, X. & Fang, Z. Quantized anomalous Hall effect in magnetic topological insulators. *Science* **329**, 61-64 (2010).

12 Chang, C. Z., Zhang, J., Feng, X., Shen, J., Zhang, Z., Guo, M., Li, K., Ou, Y., Wei, P., Wang, L. L., Ji, Z. Q., Feng, Y., Ji, S., Chen, X., Jia, J., Dai, X., Fang, Z., Zhang, S. C., He, K., Wang, Y., Lu, L., Ma, X. C. & Xue, Q. K. Experimental observation of the quantum anomalous Hall effect in a magnetic topological insulator. *Science* **340**, 167-170 (2013).

13 Checkelsky, J. G., Yoshimi, R., Tsukazaki, A., Takahashi, K. S., Kozuka, Y., Falson, J.,

Kawasaki, M. & Tokura, Y. Trajectory of the anomalous Hall effect towards the quantized state in a ferromagnetic topological insulator. *Nat. Phys.* **10**, 731-736 (2014).

14 Kou, X., Guo, S.-T., Fan, Y., Pan, L., Lang, M., Jiang, Y., Shao, Q., Nie, T., Murata, K. & Tang, J. Scale-invariant quantum anomalous Hall effect in magnetic topological insulators beyond the two-dimensional limit. *Phys. Rev. Lett.* **113**, 137201 (2014).

15 Haldane, F. D. M. Model for a quantum Hall effect without Landau levels: Condensed-matter realization of the "parity anomaly". *Phys. Rev. Lett.* **61**, 2015 (1988).

16 Xiao, D., Jiang, J., Shin, J. H., Wang, W., Wang, F., Zhao, Y. F., Liu, C., Wu, W., Chan, M. H. W., Samarth, N. & Chang, C. Z. Realization of the Axion Insulator State in Quantum Anomalous Hall Sandwich Heterostructures. *Phys. Rev. Lett.* **120**, 056801 (2018).

17 Mogi, M., Kawamura, M., Tsukazaki, A., Yoshimi, R., Takahashi, K. S., Kawasaki, M. & Tokura, Y. Tailoring tricolor structure of magnetic topological insulator for robust axion insulator. *Sci. Adv.* **3**, eaao1669 (2017).

18 Sciences, R. S. A. o. *Scientific Background on the Nobel Prize in Physics 2016: Topological Phase Transitions and Topological Phases of Matter*, https://www.nobelprize.org/nobel_prizes/physics/laureates/2016/advanced-physicsprize2016.pdf (2016).

19 Chang, C. Z., Zhao, W., Kim, D. Y., Zhang, H., Assaf, B. A., Heiman, D., Zhang, S. C., Liu, C., Chan, M. H. & Moodera, J. S. High-precision realization of robust quantum anomalous Hall state in a hard ferromagnetic topological insulator. *Nat. Mater.* **14**, 473-477 (2015).

20 Chang, C. Z., Zhao, W., Kim, D. Y., Wei, P., Jain, J. K., Liu, C., Chan, M. H. & Moodera, J. S. Zero-Field Dissipationless Chiral Edge Transport and the Nature of Dissipation in the Quantum Anomalous Hall State. *Phys. Rev. Lett.* **115**, 057206 (2015).

21 Chang, C.-Z., Zhao, W., Li, J., Jain, J., Liu, C., Moodera, J. S. & Chan, M. H. Observation of the quantum anomalous Hall insulator to Anderson insulator quantum phase transition and its scaling behavior. *Phys. Rev. Lett.* **117**, 126802 (2016).

22 Mogi, M., Yoshimi, R., Tsukazaki, A., Yasuda, K., Kozuka, Y., Takahashi, K. S., Kawasaki,

M. & Tokura, Y. Magnetic modulation doping in topological insulators toward higher-temperature quantum anomalous Hall effect. *Appl. Phys. Lett.* **107**, 182401 (2015).

23 Kandala, A., Richardella, A., Kempinger, S., Liu, C.-X. & Samarth, N. Giant anisotropic magnetoresistance in a quantum anomalous Hall insulator. *Nat. Commun.* **6**, 7434 (2015).

24 Grauer, S., Schreyeck, S., Winnerlein, M., Brunner, K., Gould, C. & Molenkamp, L. Coincidence of superparamagnetism and perfect quantization in the quantum anomalous Hall state. *Phys. Rev. B* **92**, 201304 (2015).

25 Bestwick, A., Fox, E., Kou, X., Pan, L., Wang, K. L. & Goldhaber-Gordon, D. Precise quantization of the anomalous Hall effect near zero magnetic field. *Phys. Rev. Lett.* **114**, 187201 (2015).

26 Kou, X., Pan, L., Wang, J., Fan, Y., Choi, E. S., Lee, W. L., Nie, T., Murata, K., Shao, Q., Zhang, S. C. & Wang, K. L. Metal-to-insulator switching in quantum anomalous Hall states. *Nat. Commun.* **6**, 8474 (2015).

27 Jiang, J., Xiao, D., Wang, F., Shin, J.-H., Andreoli, D., Zhang, J., Xiao, R., Zhao, Y.-F., Kayyalha, M. & Zhang, L. Concurrence of quantum anomalous Hall and topological Hall effects in magnetic topological insulator sandwich heterostructures. *Nat. Mater.* **19**, 732-737 (2020).

28 Ou, Y., Liu, C., Jiang, G., Feng, Y., Zhao, D., Wu, W., Wang, X. X., Li, W., Song, C. & Wang, L. L. Enhancing the quantum anomalous Hall effect by magnetic codoping in a topological insulator. *Adv. Mater.* **30**, 1703062 (2017).

29 Lippertz, G., Bliesener, A., Uday, A., Pereira, L. M., Taskin, A. & Ando, Y. Current-induced breakdown of the quantum anomalous Hall effect. *Phys. Rev. B* **106**, 045419 (2022).

30 Grauer, S., Fijalkowski, K., Schreyeck, S., Winnerlein, M., Brunner, K., Thomale, R., Gould, C. & Molenkamp, L. Scaling of the quantum anomalous Hall effect as an indicator of axion electrodynamics. *Phys. Rev. Lett.* **118**, 246801 (2017).

31 Fijalkowski, K., Liu, N., Hartl, M., Winnerlein, M., Mandal, P., Coschizza, A., Fothergill, A., Grauer, S., Schreyeck, S. & Brunner, K. Any axion insulator must be a bulk three-dimensional topological insulator. *Phys. Rev. B* **103**, 235111 (2021).

32 Liu, M., Wang, W., Richardella, A. R., Kandala, A., Li, J., Yazdani, A., Samarth, N. & Ong, N. P. Large discrete jumps observed in the transition between Chern states in a ferromagnetic topological insulator. *Sci. Adv.* **2**, e1600167 (2016).

33 Bai, Y., Li, Y., Luan, J., Liu, R., Song, W., Chen, Y., Ji, P.-F., Zhang, Q., Meng, F. & Tong, B. Quantized anomalous Hall resistivity achieved in molecular beam epitaxy-grown MnBi2Te4 thin films. *Natl. Sci. Rev.* **11**, nwad189 (2024).

34 Deng, Y., Yu, Y., Shi, M. Z., Guo, Z., Xu, Z., Wang, J., Chen, X. H. & Zhang, Y. Quantum anomalous Hall effect in intrinsic magnetic topological insulator $MnBi_2Te_4$. *Science* **367**, 895-900 (2020).

35 Wang, Y., Fu, B., Wang, Y., Lian, Z., Yang, S., Li, Y., Xu, L., Gao, Z., Yang, X. & Wang, W. Towards the quantized anomalous Hall effect in $AlO_x$-capped $MnBi_2Te_4$. *Nat. Commun.* **16**, 1727 (2025).

36 Lian, Z., Wang, Y., Wang, Y., Dong, W.-H., Feng, Y., Dong, Z., Ma, M., Yang, S., Xu, L. & Li, Y. Antiferromagnetic quantum anomalous Hall effect under spin flips and flops. *Nature* **641**, 70-75 (2025).

37 Lu, Z., Han, T., Yao, Y., Reddy, A. P., Yang, J., Seo, J., Watanabe, K., Taniguchi, T., Fu, L. & Ju, L. Fractional quantum anomalous Hall effect in multilayer graphene. *Nature* **626**, 759-764 (2024).

38 Han, T., Lu, Z., Yao, Y., Yang, J., Seo, J., Yoon, C., Watanabe, K., Taniguchi, T., Fu, L. & Zhang, F. Large quantum anomalous Hall effect in spin-orbit proximitized rhombohedral graphene. *Science* **384**, 647-651 (2024).

39 Xie, J., Huo, Z., Lu, X., Feng, Z., Zhang, Z., Wang, W., Yang, Q., Watanabe, K., Taniguchi, T. & Liu, K. Tunable fractional Chern insulators in rhombohedral graphene superlattices. *Nat. Mater.* **24**, 1042-1048 (2025).

40 Serlin, M., Tschirhart, C., Polshyn, H., Zhang, Y., Zhu, J., Watanabe, K., Taniguchi, T., Balents, L. & Young, A. Intrinsic quantized anomalous Hall effect in a moiré heterostructure.

*Science* **367**, 900-903 (2020).

41 Li, T., Jiang, S., Shen, B., Zhang, Y., Li, L., Tao, Z., Devakul, T., Watanabe, K., Taniguchi, T. & Fu, L. Quantum anomalous Hall effect from intertwined moiré bands. *Nature* **600**, 641-646 (2021).

42 Park, H., Cai, J., Anderson, E., Zhang, Y., Zhu, J., Liu, X., Wang, C., Holtzmann, W., Hu, C. & Liu, Z. Observation of fractionally quantized anomalous Hall effect. *Nature* **622**, 74-79 (2023).

43 Xu, F., Sun, Z., Jia, T., Liu, C., Xu, C., Li, C., Gu, Y., Watanabe, K., Taniguchi, T. & Tong, B. Observation of integer and fractional quantum anomalous Hall effects in twisted bilayer $MoTe_2$. *Phys. Rev. X* **13**, 031037 (2023).

44 Zhao, Y. F., Zhang, R., Mei, R., Zhou, L. J., Yi, H., Zhang, Y. Q., Yu, J., Xiao, R., Wang, K., Samarth, N., Chan, M. H. W., Liu, C. X. & Chang, C. Z. Tuning the Chern number in quantum anomalous Hall insulators. *Nature* **588**, 419-423 (2020).

45 Li, Y.-H. & Cheng, R. Spin fluctuations in quantized transport of magnetic topological insulators. *Phys. Rev. Lett.* **126**, 026601 (2021).

46 Lin, W., Feng, Y., Wang, Y., Lian, Z., Li, H., Wu, Y., Liu, C., Wang, Y., Zhang, J. & Wang, Y. Influence of the dissipative topological edge state on quantized transport in $MnBi_2Te_4$. *Phys. Rev. B* **105**, 165411 (2022).

47 Lachman, E. O., Young, A. F., Richardella, A., Cuppens, J., Naren, H., Anahory, Y., Meltzer, A. Y., Kandala, A., Kempinger, S. & Myasoedov, Y. Visualization of superparamagnetic dynamics in magnetic topological insulators. *Sci. Adv.* **1**, e1500740 (2015).

48 Feng, X., Feng, Y., Wang, J., Ou, Y., Hao, Z., Liu, C., Zhang, Z., Zhang, L., Lin, C. & Liao, J. Thickness Dependence of the Quantum Anomalous Hall Effect in Magnetic Topological Insulator Films. *Adv. Mater.* **28**, 6386-6390 (2016).

49 Kawamura, M., Yoshimi, R., Tsukazaki, A., Takahashi, K. S., Kawasaki, M. & Tokura, Y. Current-driven instability of the quantum anomalous Hall effect in ferromagnetic topological insulators. *Phys. Rev. Lett.* **119**, 016803 (2017).

50 Li, W., Claassen, M., Chang, C. Z., Moritz, B., Jia, T., Zhang, C., Rebec, S., Lee, J. J., Hashimoto, M., Lu, D. H., Moore, R. G., Moodera, J. S., Devereaux, T. P. & Shen, Z. X. Origin of the low critical observing temperature of the quantum anomalous Hall effect in V-doped $(Bi, Sb)_2Te_3$ film. *Sci. Rep.* **6**, 32732 (2016).

51 Wang, W., Ou, Y., Liu, C., Wang, Y., He, K., Xue, Q.-K. & Wu, W. Direct evidence of ferromagnetism in a quantum anomalous Hall system. *Nat. Phys.* **14**, 791-795 (2018).

52 Zhao, Y. F., Zhang, R., Sun, Z. T., Zhou, L. J., Zhuo, D., Yan, Z. J., Yi, H., Wang, K., Chan, M. H. & Liu, C. X. 3D Quantum Anomalous Hall Effect in Magnetic Topological Insulator Trilayers of Hundred-Nanometer Thickness. *Adv. Mater.* **36**, 2310249 (2024).

53 Wilczek, F. Two applications of axion electrodynamics. *Phys. Rev. Lett.* **58**, 1799 (1987).

54 Fang, C., Gilbert, M. J. & Bernevig, B. A. Large-Chern-number quantum anomalous Hall effect in thin-film topological crystalline insulators. *Phys. Rev. Lett.* **112**, 046801 (2014).

55 Wang, J., Lian, B., Zhang, H., Xu, Y. & Zhang, S.-C. Quantum anomalous Hall effect with higher plateaus. *Phys. Rev. Lett.* **111**, 136801 (2013).

56 Jiang, H., Qiao, Z., Liu, H. & Niu, Q. Quantum anomalous Hall effect with tunable Chern number in magnetic topological insulator film. *Phys. Rev. B* **85**, 045445 (2012).

57 Zhang, J., Chang, C.-Z., Tang, P., Zhang, Z., Feng, X., Li, K., Wang, L.-l., Chen, X., Liu, C. & Duan, W. Topology-driven magnetic quantum phase transition in topological insulators. *Science* **339**, 1582-1586 (2013).

58 Chang, C.-Z., Tang, P., Wang, Y.-L., Feng, X., Li, K., Zhang, Z., Wang, Y., Wang, L.-L., Chen, X. & Liu, C. Chemical-potential-dependent gap opening at the Dirac surface states of $Bi_2Se_3$ induced by aggregated substitutional Cr atoms. *Phys. Rev. Lett.* **112**, 056801 (2014).

59 Wang, F., Zhang, H., Jiang, J., Zhao, Y.-F., Yu, J., Liu, W., Li, D., Chan, M. H., Sun, J. & Zhang, Z. Chromium-induced ferromagnetism with perpendicular anisotropy in topological crystalline insulator SnTe (111) thin films. *Phys. Rev. B* **97**, 115414 (2018).

60 Burkov, A. & Balents, L. Weyl semimetal in a topological insulator multilayer. *Phys. Rev. Lett.* **107**, 127205 (2011).

61 Lei, C., Chen, S. & MacDonald, A. H. Magnetized topological insulator multilayers. *Proc. Natl. Acad. Sci.* **117**, 27224-27230 (2020).

62 Jiang, G., Feng, Y., Wu, W., Li, S., Bai, Y., Li, Y., Zhang, Q., Gu, L., Feng, X. & Zhang, D. Quantum anomalous Hall multilayers grown by molecular beam epitaxy. *Chin. Phys. Lett.* **35**, 076802 (2018).

63 Zhao, Y.-F., Zhang, R., Zhou, L.-J., Mei, R., Yan, Z.-J., Chan, M. H., Liu, C.-X. & Chang, C.-Z. Zero magnetic field plateau phase transition in higher chern number quantum anomalous hall insulators. *Phys. Rev. Lett.* **128**, 216801 (2022).

64 Zhou, L.-J., Zhuo, D., Mei, R., Zhao, Y.-F., Yang, K., Zhang, R., Yan, Z., Tay, H., Chan, M. H. & Liu, C.-X. Interlayer coupling induced quantum phase transition in quantum anomalous Hall multilayers. *Phys. Rev. B* **111**, L201304 (2025).

65 Pruisken, A. Universal singularities in the integral quantum Hall effect. *Phys. Rev. Lett.* **61**, 1297 (1988).

66 Wei, H., Tsui, D., Paalanen, M. & Pruisken, A. Experiments on delocalization and university in the integral quantum Hall effect. *Phys. Rev. Lett.* **61**, 1294 (1988).

67 Li, W., Csáthy, G., Tsui, D., Pfeiffer, L. & West, K. Scaling and universality of integer quantum Hall plateau-to-plateau transitions. *Phys. Rev. Lett.* **94**, 206807 (2005).

68 Li, W., Vicente, C., Xia, J., Pan, W., Tsui, D., Pfeiffer, L. & West, K. Scaling in Plateau-to-Plateau Transition: A Direct Connection of Quantum Hall Systems with the Anderson Localization Model. *Phys. Rev. Lett.* **102**, 216801 (2009).

69 Zhang, J. in *Transport Studies of the Electrical, Magnetic and Thermoelectric properties of Topological Insulator Thin Films* 55-86 (Springer, 2016).

70 Liu, C.-X., Qi, X.-L., Dai, X., Fang, Z. & Zhang, S.-C. Quantum anomalous Hall effect in $Hg_{1-y}Mn_yTe$ quantum wells. *Phys. Rev. Lett.* **101**, 146802 (2008).

71 Zhuo, D., Zhou, L., Zhao, Y.-F., Zhang, R., Yan, Z.-J., Wang, A. G., Chan, M. H., Liu, C.-X., Chen, C.-Z. & Chang, C.-Z. Engineering Plateau Phase Transition in Quantum Anomalous Hall Multilayers. *Nano. Lett.* **24**, 6974-6980 (2024).

72 Deng, P., Han, Y., Zhang, P., Chong, S. K., Qiao, Z. & Wang, K. L. Tuning the number of chiral edge channels in a fixed quantum anomalous Hall system. *Phys. Rev. B* **109**, L201402 (2024).

73 Hilke, M., Shahar, D., Song, S. H., Tsui, D. C., Xie, Y. H. & Shayegan, M. Semicircle: An exact relation in the integer and fractional quantum Hall effect. *Europhys. Lett.* **46**, 775-779 (1999).

74 Fox, E. J., Rosen, I. T., Yang, Y., Jones, G. R., Elmquist, R. E., Kou, X., Pan, L., Wang, K. L. & Goldhaber-Gordon, D. Part-per-million quantization and current-induced breakdown of the quantum anomalous Hall effect. *Phys. Rev. B* **98**, 075145 (2018).

75 Yasuda, K., Mogi, M., Yoshimi, R., Tsukazaki, A., Takahashi, K., Kawasaki, M., Kagawa, F. & Tokura, Y. Quantized chiral edge conduction on domain walls of a magnetic topological insulator. *Science* **358**, 1311-1314 (2017).

76 Wang, S.-W., Xiao, D., Dou, Z., Cao, M., Zhao, Y.-F., Samarth, N., Chang, C.-Z., Connolly, M. R. & Smith, C. G. Demonstration of dissipative quasihelical edge transport in quantum anomalous hall insulators. *Phys. Rev. Lett.* **125**, 126801 (2020).

77 Kawamura, M., Mogi, M., Yoshimi, R., Tsukazaki, A., Kozuka, Y., Takahashi, K. S., Kawasaki, M. & Tokura, Y. Current scaling of the topological quantum phase transition between a quantum anomalous Hall insulator and a trivial insulator. *Phys. Rev. B* **102**, 041301 (2020).

78 Chen, C.-Z., Xie, Y.-M., Liu, J., Lee, P. A. & Law, K. T. Quasi-one-dimensional quantum anomalous Hall systems as new platforms for scalable topological quantum computation. *Phys. Rev. B* **97**, 104504 (2018).

79 Qiu, G., Zhang, P., Deng, P., Chong, S. K., Tai, L., Eckberg, C. & Wang, K. L. Mesoscopic transport of quantum anomalous Hall effect in the submicron size regime. *Phys. Rev. Lett.* **128**, 217704 (2022).

80 Fijalkowski, K. M., Liu, N., Mandal, P., Schreyeck, S., Brunner, K., Gould, C. & Molenkamp, L. W. Macroscopic quantum tunneling of a topological ferromagnet. *Adv. Sci.* **10**, 2303165

(2023).

81 Zhou, L. J., Mei, R., Zhao, Y. F., Zhang, R., Zhuo, D., Yan, Z. J., Yuan, W., Kayyalha, M., Chan, M. H. W., Liu, C. X. & Chang, C. Z. Confinement-Induced Chiral Edge Channel Interaction in Quantum Anomalous Hall Insulators. *Phys. Rev. Lett.* **130**, 086201 (2023).

82 Manchon, A., Železný, J., Miron, I. M., Jungwirth, T., Sinova, J., Thiaville, A., Garello, K. & Gambardella, P. Current-induced spin-orbit torques in ferromagnetic and antiferromagnetic systems. *Rev. Mod. Phys.* **91**, 035004 (2019).

83 Yuan, W., Zhou, L.-J., Yang, K., Zhao, Y.-F., Zhang, R., Yan, Z., Zhuo, D., Mei, R., Wang, Y. & Yi, H. Electrical switching of the edge current chirality in quantum anomalous Hall insulators. *Nat. Mater.* **23**, 58-64 (2024).

84 Morimoto, T., Furusaki, A. & Nagaosa, N. Topological magnetoelectric effects in thin films of topological insulators. *Phys. Rev. B* **92**, 085113 (2015).

85 Wang, J., Lian, B., Qi, X.-L. & Zhang, S.-C. Quantized topological magnetoelectric effect of the zero-plateau quantum anomalous Hall state. *Phys. Rev. B* **92**, 081107 (2015).

86 Mogi, M., Kawamura, M., Yoshimi, R., Tsukazaki, A., Kozuka, Y., Shirakawa, N., Takahashi, K., Kawasaki, M. & Tokura, Y. A magnetic heterostructure of topological insulators as a candidate for an axion insulator. *Nat. Mater.* **16**, 516-521 (2017).

87 Lachman, E. O., Mogi, M., Sarkar, J., Uri, A., Bagani, K., Anahory, Y., Myasoedov, Y., Huber, M. E., Tsukazaki, A. & Kawasaki, M. Observation of superparamagnetism in coexistence with quantum anomalous Hall $C = \pm 1$ and $C = 0$ Chern states. *npj Quantum Mater.* **2**, 70 (2017).

88 Zhuo, D., Yan, Z.-J., Sun, Z. T., Zhou, L.-J., Zhao, Y.-F., Zhang, R., Mei, R., Yi, H., Wang, K., Chan, M. H. W., Liu, C.-X., Law, K. T. & Chang, C.-Z. Axion insulator state in hundred-nanometer-thick magnetic topological insulator sandwich heterostructures. *Nat. Commun.* **14**, 7596 (2023).

89 Wu, X., Xiao, D., Chen, C.-Z., Sun, J., Zhang, L., Chan, M. H. W., Samarth, N., Xie, X. C., Lin, X. & Chang, C.-Z. Scaling behavior of the quantum phase transition from a quantum-anomalous-Hall insulator to an axion insulator. *Nat. Commun.* **11**, 4532 (2020).

90 Feng, Y., Feng, X., Ou, Y., Wang, J., Liu, C., Zhang, L., Zhao, D., Jiang, G., Zhang, S. C., He, K., Ma, X., Xue, Q. K. & Wang, Y. Observation of the Zero Hall Plateau in a Quantum Anomalous Hall Insulator. *Phys. Rev. Lett.* **115**, 126801 (2015).

91 Murakami, S. Phase transition between the quantum spin Hall and insulator phases in 3D: emergence of a topological gapless phase. *New J. Phys.* **9**, 356 (2007).

92 Yang, B.-J. & Nagaosa, N. Classification of stable three-dimensional Dirac semimetals with nontrivial topology. *Nat. Commun.* **5**, 4898 (2014).

93 Armitage, N., Mele, E. & Vishwanath, A. Weyl and Dirac semimetals in three-dimensional solids. *Rev. Mod. Phys.* **90**, 015001 (2018).

94 Niemi, A. J. & Semenoff, G. W. Axial-anomaly-induced fermion fractionization and effective gauge-theory actions in odd-dimensional space-times. *Phys. Rev. Lett.* **51**, 2077 (1983).

95 Jackiw, R. Fractional charge and zero modes for planar systems in a magnetic field. *Phys. Rev. D* **29**, 2375 (1984).

96 Redlich, A. N. Gauge noninvariance and parity nonconservation of three-dimensional fermions. *Phys. Rev. Lett.* **52**, 18 (1984).

97 Boyanovsky, D., Blankenbecler, R. & Yahalom, R. Physical origin of topological mass in 2+1 dimensions. *Nucl. Phys. B.* **270**, 483-505 (1986).

98 Schakel, A. M. Relativistic quantum Hall effect. *Phys. Rev. D* **43**, 1428 (1991).

99 Semenoff, G. W. Condensed-matter simulation of a three-dimensional anomaly. *Phys. Rev. Lett.* **53**, 2449 (1984).

100 Fradkin, E., Dagotto, E. & Boyanovsky, D. Physical realization of the parity anomaly in condensed matter physics. *Phys. Rev. Lett.* **57**, 2967 (1986).

101 Lu, R., Sun, H., Kumar, S., Wang, Y., Gu, M., Zeng, M., Hao, Y.-J., Li, J., Shao, J. & Ma, X.-M. Half-magnetic topological insulator with magnetization-induced Dirac gap at a selected surface. *Phys. Rev. X* **11**, 011039 (2021).

102 Mogi, M., Okamura, Y., Kawamura, M., Yoshimi, R., Yasuda, K., Tsukazaki, A., Takahashi, K., Morimoto, T., Nagaosa, N. & Kawasaki, M. Experimental signature of the parity anomaly in a semi-magnetic topological insulator. *Nat. Phys.* **18**, 390-394 (2022).

103 Zhuo, D., Zhang, B., Zhou, H., Tay, H., Liu, X., Xi, Z., Chen, C.-Z. & Chang, C.-Z. Half-quantized chiral edge current in a $C = 1/2$ parity anomaly state. *Phys. Rev. Lett.* **136**, 016601 (2026).

104 Wang, B., Hu, J., Fu, B., Li, J., Kong, Y., Bai, K.-Z., Shen, S.-Q. & Xiao, D. Parity Anomalous Semimetal with Minimal Conductivity Induced by an In-Plane Magnetic Field. *Phys. Rev. Lett.* **136**, 146602 (2026).

105 Hu, J., Wang, B., Zhou, H., Jia, T., Sun, Z., Liu, C., Zhang, B., Qian, D., Li, T. & Xie, X. Half-quantized layer hall effect as a probe of quantized axion field. *Nat. Commun.* (2026).

106 Yang, T. H., Li, Y., Zhang, P., Yao, Y. T., Yang, H. Y., Shu, Q., Choi, E. S., Wong, K., Chang, T. R. & Qiu, G. In-Plane Field Induced Half Quantized Hall Conductivity in Trilayer Magnetic Topological Insulator. *Adv. Mater.* **38**, e10754 (2026).

107 Chen, Y., Chu, J.-H., Analytis, J., Liu, Z., Igarashi, K., Kuo, H.-H., Qi, X., Mo, S.-K., Moore, R. & Lu, D. Massive Dirac fermion on the surface of a magnetically doped topological insulator. *Science* **329**, 659-662 (2010).

108 Yoshimi, R., Yasuda, K., Tsukazaki, A., Takahashi, K., Nagaosa, N., Kawasaki, M. & Tokura, Y. Quantum Hall states stabilized in semi-magnetic bilayers of topological insulators. *Nat. Commun.* **6**, 8530 (2015).

109 Zhou, H., Li, H., Xu, D.-H., Chen, C.-Z., Sun, Q.-F. & Xie, X. Transport theory of half-quantized Hall conductance in a semimagnetic topological insulator. *Phys. Rev. Lett.* **129**, 096601 (2022).

110 Chu, R.-L., Shi, J. & Shen, S.-Q. Surface edge state and half-quantized Hall conductance in topological insulators. *Phys. Rev. B* **84**, 085312 (2011).

111 Zou, J.-Y., Fu, B., Wang, H.-W., Hu, Z.-A. & Shen, S.-Q. Half-quantized Hall effect and power law decay of edge-current distribution. *Phys. Rev. B* **105**, L201106 (2022).

112 Wan, Y.-H., Liu, P.-Y. & Sun, Q.-F. Quantum anomalous Hall effect in ferromagnetic metals. *Phys. Rev. Lett.* **135**, 186302 (2025).

113 Bai, K.-Z., Fu, B., Zhang, Z. & Shen, S.-Q. Metallic quantized anomalous Hall effect without chiral edge states. *Phys. Rev. B* **108**, L241407 (2023).

114 Bi, S.-H., Fu, B. & Shen, S.-Q. Half-quantized Hall metal and marginal metal in disordered magnetic topological insulators. *Communications Physics* **8**, 332 (2025).

115 Zhou, H., Chen, C.-Z., Sun, Q.-F. & Xie, X. Dissipative chiral channels, Ohmic scaling, and half-integer Hall conductivity from relativistic quantum Hall effect. *Phys. Rev. B* **109**, 115305 (2024).

116 Burkov, A., Hook, M. & Balents, L. Topological nodal semimetals. *Phys. Rev. B* **84**, 235126 (2011).

117 Nomura, K. & Nagaosa, N. Surface-Quantized Anomalous Hall Current and the Magnetoelectric Effect in Magnetically Disordered Topological Insulators. *Phys. Rev. Lett.* **106**, 166802 (2011).

118 Essin, A. M., Moore, J. E. & Vanderbilt, D. Magnetoelectric polarizability and axion electrodynamics in crystalline insulators. *Phys. Rev. Lett.* **102**, 146805 (2009).

119 Tse, W.-K. & MacDonald, A. H. Giant Magneto-Optical Kerr Effect and Universal Faraday Effect in Thin-Film Topological Insulators. *Phys. Rev. Lett.* **105**, 057401 (2010).

120 Maciejko, J., Qi, X.-L., Drew, H. D. & Zhang, S.-C. Topological quantization in units of the fine structure constant. *Phys. Rev. Lett.* **105**, 166803 (2010).

121 Qi, X.-L., Li, R., Zang, J. & Zhang, S.-C. Inducing a magnetic monopole with topological surface states. *Science* **323**, 1184-1187 (2009).